\documentclass[twocolumn,showpacs,preprintnumbers,linenumbers,amsmath,amssymb]{revtex4-1}

\usepackage{graphicx}
\usepackage{color}

\usepackage{footnote}
\makeatletter

\begin{document}
\widetext

\title{\boldmath Study of $J/\psi \rightarrow \omega p \bar{p}$ at BESIII}

\date{\today}

\author{
\small
M.~Ablikim$^{1}$, M.~N.~Achasov$^{6}$, O.~Albayrak$^{3}$, D.~J.~Ambrose$^{39}$, F.~F.~An$^{1}$, Q.~An$^{40}$, J.~Z.~Bai$^{1}$, R.~Baldini Ferroli$^{17A}$, Y.~Ban$^{26}$, J.~Becker$^{2}$
% ...
\\
\vspace{0.2cm}
(BESIII Collaboration)\\
\vspace{0.2cm} {\it
$^{1}$ Institute of High Energy Physics, Beijing 100049, People's Republic of China\\
$^{2}$ Bochum Ruhr-University, D-44780 Bochum, Germany\\
$^{3}$ Carnegie Mellon University, Pittsburgh, Pennsylvania 15213, USA\\
% ...
$^{6}$ G.I. Budker Institute of Nuclear Physics SB RAS (BINP), Novosibirsk 630090, Russia\\
% ...
$^{17}$ (A)INFN Laboratori Nazionali di Frascati, I-00044, Frascati, Italy; (B)INFN and University of Perugia, I-06100, Perugia, Italy\\
% ...
$^{26}$ Peking University, Beijing 100871, People's Republic of China\\
% ...
$^{39}$ University of Rochester, Rochester, New York 14627, USA\\
$^{40}$ University of Science and Technology of China, Hefei 230026, People's Republic of China\\
% ...
}}

\begin{abstract}
The decay $J/\psi \rightarrow \omega p \bar{p}$ has been
studied, using $225.3\times 10^{6}$ $J/\psi$ events accumulated at
BESIII. No significant enhancement near the $p\bar{p}$ invariant-mass
threshold (denoted as $X(p\bar{p})$) is observed. The upper limit of the branching fraction
$\mathcal{B}(J/\psi \rightarrow \omega X(p\bar{p}) \rightarrow \omega p \bar{p})$ is
determined to be $3.9\times10^{-6}$ at the $95\%$ confidence level.
The branching fraction of  $J/\psi \rightarrow \omega p \bar{p}$ is
measured to be $\mathcal{B}(J/\psi \rightarrow \omega p \bar{p})
=(9.0 \pm 0.2\ (\text{stat.})\pm 0.9\ (\text{syst.})) \times
10^{-4}$.

\end{abstract}


\maketitle
\righthyphenmin=2
\lefthyphenmin=2

\section{Introduction}
An anomalous enhancement near the threshold of the $p\bar{p}$ system,
namely $X(p\bar{p})$, was first observed by the BESII experiment in
the radiative decay $J/\psi\rightarrow \gamma p \bar{p}$
~\cite{BESII_gppbar}, and it was recently confirmed by the CLEO and BESIII
experiments~\cite{BAM_00004,BAM_00004b,CLEO2010}. In the BESII
experiment, its mass is measured to be $1859^{+3}_{-10} \
(\text{stat.})^{+5}_{-25} \ (\text{syst.})\, \text{MeV}/c^2$ and
the total width is $\Gamma < 30\, \text{MeV}/c^2$ at the 90\%
confidence level (C.L.). While in the BESIII experiment, a partial
wave analysis (PWA) with a correction for the final-state interaction (FSI) is
performed, and the spin-parity of $X(p\bar{p})$ is determined to be
$0^{-+}$, its mass is $1832^{+19}_{-5}\ (\text{stat.})^{+18}_{-17}\
(\text{syst.})\,\text{MeV}/c^{2}$ and the total width is $\Gamma <
76 \, \text{MeV}/c^2$ at the 90\% C.L. ~\cite{BAM_00004b}.

The discovery of $X(p\bar{p})$ stimulated a number of theoretical interpretations
and experimental studies~\cite{CLEO2006,BESII_omegappbar,
BESII_psi_rediative,Ex_Rev,Datta,Yan,Zhu,Abud,Zou,Bugg,Sibirtsev,Ph_Rev}.
There is no experimental evidence of such an enhancement in
other quarkonium decays, e.g. $J/\psi \rightarrow \pi^{0}
p\bar{p}$~\cite{BESII_gppbar} or $\Upsilon(2S) \rightarrow \gamma p
\bar{p}$~\cite{CLEO2006}. In $\psi(2S)\rightarrow \gamma p\bar{p}$,
the recent $\text{BESIII}$ measurement shows a relative production rate to that of
$J/\psi$ decays of $R=5.08\%$~\cite{BAM_00004b}.
A number of theoretical speculations have been proposed to interpret the
nature of this structure, including baryonium~\cite{Datta,Yan,Zhu}, a multi-quark
state~\cite{Abud} or mainly a pure FSI~\cite{Zou,Sibirtsev}. It was proposed
to associate this enhancement with a broad enhancement observed
in $B$ meson decays~\cite{BELLE,BABAR} or a new resonance $X(1835)$ in
$J/\psi \rightarrow \gamma \pi^{+}\pi^{-}\eta^{'}$ decay at BESII~\cite{BESII_X1835}.

The investigation of the near-threshold $p\bar{p}$ invariant mass
spectrum in other $J/\psi$ decay modes will be helpful in
understanding the nature of the observed structure.  The decay $J/\psi
\rightarrow \omega p\bar{p}$ restricts the isospin of the $p\bar{p}$
system, and it is helpful to clarify the role of the $\ p\bar{p}$
FSI. The BESII collaboration studied $J/\psi \rightarrow \omega p
\bar{p}$ via $\omega$ decaying to $\pi^{0}\pi^{+}\pi^{-}$ with a data
sample of $5.8\times10^{7}$ $J/\psi$
events~\cite{BESII_omegappbar}. No significant signal near the
threshold of the $p\bar{p}$ invariant-mass spectrum was observed and
an upper limit on the branching fraction of $J/\psi\rightarrow \omega
X(p\bar{p}) \rightarrow \omega p \bar{p}$ was determined to be
$1.5\times 10^{-5}$ at the 90\% C.L., which disfavored the
interpretation of a pure FSI effect giving rise to the
$X(p\bar{p})$. In this paper, the analysis of $J/\psi\rightarrow
\omega p\bar{p}$ via the decay channel $\omega \rightarrow \gamma
\pi^{0}$ is presented, based on a data sample of $(225.3\pm 2.8)
\times10^{6}$ $J/\psi$ events~\cite{BAM_00011} accumulated with the
BESIII detector. Searching for the $X(p\bar{p})$ in the decay mode
$J/\psi \rightarrow \omega p \bar{p} \rightarrow \gamma \pi^0
p\bar{p}$ has a particular advantage: a low irreducible background
from $N^*$ is expected. The channel $J/\psi \rightarrow \omega
p\bar{p}\rightarrow \pi\pi\pi^0p\bar{p}$ has
irreducible background from various $N^{*}$ decays and $\Delta$ decays,
where interferences may have a large impact on the uncertainty of the
measurements.

BESIII/BEPCII~\cite{bes3} is a major upgrade of the $\text{BESII}$ experiment
at the BEPC accelerator~\cite{bepc} for studies of hadron spectroscopy
and $\tau$-charm physics~\cite{bes3yellow}. The design peak luminosity
of the double-ring $e^{+}e^{-}$ collider, BEPCII, is $10^{33}$\,cm$^{-2}$s$^{-1}$
at beam currents of 0.93\,A. The BESIII detector
with a geometrical acceptance of 93\% of 4$\pi$, consists of the
following main components: 1) a small-celled, helium-based main drift
chamber (MDC) with 43 layers. The average single wire resolution is
135\,$\mu$m, and the momentum resolution for 1\,GeV/$c^2$ charged particles
in a 1\,T magnetic field is 0.5\%; 2) an electromagnetic calorimeter
(EMC) made of 6240 CsI\,(Tl) crystals arranged in a cylindrical shape
(barrel) plus two end-caps. For 1.0\,GeV photons, the energy resolution
is 2.5\% in the barrel and 5\% in the end-caps, and the position
resolution is 6\,mm in the barrel and 9\,mm in the end-caps; 3) a
Time-Of-Flight system (TOF) for particle identification (PID) composed
of a barrel part made of two layers with 88 pieces of 5\,cm thick, 2.4\,m
long plastic scintillators in each layer, and two end-caps with 48
fan-shaped, 5\,cm thick, plastic scintillators in each end-cap. The time
resolution is 80\,ps in the barrel, and 110 ps in the end-caps,
corresponding to a $K/\pi$ separation by more than 2$\sigma$ for
momenta below about 1\,GeV/$c^2$; 4) a muon chamber system (MUC) made of
1000\,m$^2$ of Resistive Plate Chambers (RPC) arranged in 9 layers in
the barrel and 8 layers in the end-caps and incorporated in the return
iron yoke of the superconducting magnet.  The position resolution is
about 2\,cm.

The optimization of the event selection and the estimate of physics
backgrounds are performed through Monte Carlo (MC) simulations. The
GEANT4-based simulation software BOOST~\cite{boost} includes the
geometric and material description of the BESIII detectors and the
detector response and digitization models, as well as the tracking of
the detector running conditions and performance. The production of the
$J/\psi$ resonance is simulated by the MC event generator
KKMC~\cite{kkmc}, while the decays are generated by
EVTGEN~\cite{evtgen} for known decay modes with branching ratios being
set to PDG~\cite{pdg} world average values, and by
LUNDCHARM~\cite{lundcharm} for the remaining unknown decays. The
analysis is performed in the framework of the BESIII offline software
system~\cite{boss} which takes care of the detector calibration, event
reconstruction and data storage.

\section{Event Selection}
Signal $J/\psi \rightarrow \omega p \bar{p}$ events with $\omega\rightarrow \gamma \pi^{0}$
final states have the topology $\gamma\gamma\gamma p\bar{p}$. The event candidates are
required to have two well reconstructed charged tracks with net charge zero, and at least
three photons.

Charged-particle tracks in the polar angle range $|\cos \theta| <0.93$
are reconstructed from the MDC hits, only tracks in barrel region
($|\cos\theta|<0.8$) are used to reduce systematic uncertainties in
tracking and particle identification. Tracks with their points of
closest approach to the beamline within $\pm10\,$cm of the interaction
point in the beam direction, and within $1\,$cm in the plane
perpendicular to the beam are selected.  TOF and
$\text{d}E\text{/}\text{d}x$ information are combined to determine
particle identification confidence levels for $\pi$, $K$ and $p
(\bar{p})$ hypotheses; and the particle type with highest confidence
level is assigned to each track.  A proton and an anti-proton are
required. To reduce the systematic error due to differences of the
tracking efficiency at low momentum between data and MC, the momentum
of the proton or anti-proton is further required to be larger than
$300\,\text{MeV}/c$.

Photon candidates are reconstructed by clustering signals in EMC crystals.
The photon candidates are required to be in the barrel region ($|\cos \theta| <0.8$)
of the EMC with at least $25\, \text{MeV}$ energy deposition, or
in the end-caps region ($0.86<|\cos \theta| <0.92$) with at least $50\, \text{MeV}$
energy deposition, where $\theta$ is the polar angle of the shower.
Timing information from the EMC is used to suppress electronic noise and energy depositions
that are unrelated to the event. To suppress showers generated by charged particles,
the photon candidates are furthermore required to be separated by an angle larger than $10^{\circ}$
and larger than $30^{\circ}$ from the proton and anti-proton, respectively.

A four-constraint (4C) energy-momentum conserving kinematic fit is performed to the
$\gamma \gamma \gamma p \bar{p}$ hypothesis. For events with more than three photon candidates, the
combination with the minimum $\chi^{2}_{4C}$  is selected, and $\chi^{2}_{4C}<30$
is required. The $\pi^{0}$ candidates are reconstructed from the two of the three selected
photons with an invariant mass closest to the $\pi^{0}$ mass, and
$|M_{\gamma\gamma}-M_{\pi^{0}}|<15 \,\text{MeV}/c^2$ is required.

\section{Branching Fraction and Yield Measurements}
\begin{center}
\begin{figure}[hbt!]
\includegraphics[width=8cm]{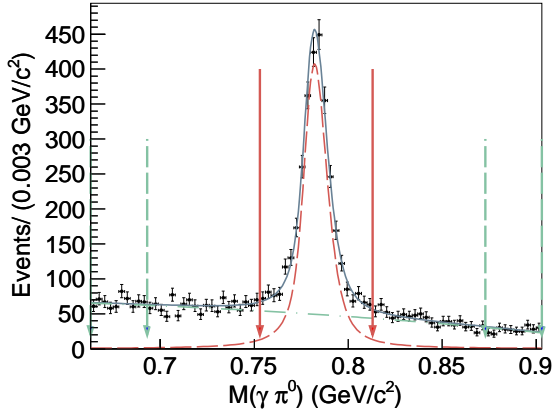}
\caption{$\gamma \pi^{0}$ invariant mass distribution of $J/\psi
\rightarrow \gamma \pi^{0} p \bar{p}$ candidates. The dashed line is
the signal shape which is parametrized by a Breit-Wigner function
convoluted with the detector resolution described by the Novosibirsk
function; the dashed-dotted line is the background shape which is
described by a second order Chebychev polynomial; and the solid line is
the total contribution of the two components.  The solid arrows
indicate the $\omega$ signal region \,($0.753<M(\gamma
\pi^{0})<0.813\,\text{GeV}/c^{2}$) and the two pairs of dashed arrows
indicate the $\omega$ sidebands\,($0.663<M(\gamma
\pi^{0})<0.693\,\text{GeV}/c^{2}$ and $0.873<M(\gamma
\pi^{0})<0.903\,\text{GeV}/c^{2}$).  }
\label{omegafit}
\end{figure}
\end{center}

Figure~1 shows the $\gamma \pi^{0}$ invariant mass spectrum for
candidate $J/\psi \rightarrow \gamma \pi^{0} p \bar{p}$ events, where
a distinctive $\omega$ signal is seen.  An unbinned maximum likelihood
fit is performed to the $\gamma \pi^{0}$ invariant mass with the
$\omega$ signal parametrized by a Breit-Wigner function convoluted
with the Novosibirsk function~\cite{roofit} which describes the
detector resolution.  The background shape is described by a
second-order Chebychev polynomial function.  The mass and width of the
$\omega$ peak are fixed to the values published by the Particle Data
Group (PDG)~\cite{pdg}, and the yield of the $\omega$ signal obtained
from the fit is $N_{obs}=2670\pm69$.

The branching fraction of $J/\psi \rightarrow \omega p\bar{p}$ is calculated according to :
\begin{equation}
\mathcal{B}(J/\psi\rightarrow \omega p\bar{p}) =
\frac{N_{\text{obs}}}{N_{J/\psi}\times
\mathcal{B}(\omega\rightarrow \gamma \pi^{0})\times \mathcal{B}(\pi^{0}\rightarrow
\gamma\gamma)\times \varepsilon_{\text{rec}}}.
\end{equation}
where $N_{\text{obs}}$ is the number of signal events determined from
the fit to the $\gamma\pi^0$ invariant mass; $N_{J/\psi}$ is the
number of $J/\psi$ events~\cite{BAM_00011}; $\mathcal{B}(\omega
\rightarrow \gamma \pi^{0})$ and $\mathcal{B}(\pi^{0} \rightarrow
\gamma\gamma)$ are branching fractions of $\omega \rightarrow \gamma
\pi^{0}$ and $\pi^{0} \rightarrow \gamma\gamma$, respectively, as
from the PDG~\cite{pdg}; and the detection efficiency
$\varepsilon_{\text{rec}}$ is $(16.1\pm 1.7)\%$ obtained from a MC
sample for $J/\psi \rightarrow \omega p \bar{p}$ events generated
according to a phase-space distribution. The measured branching
fraction is $\mathcal{B}( J/\psi\rightarrow \omega p\bar{p})$ $=$
$(9.0\pm0.2 \, (\text{stat.}))\times 10^{-4} $.

\begin{widetext}

\begin{figure}[hbt!]

\includegraphics[width=16cm]{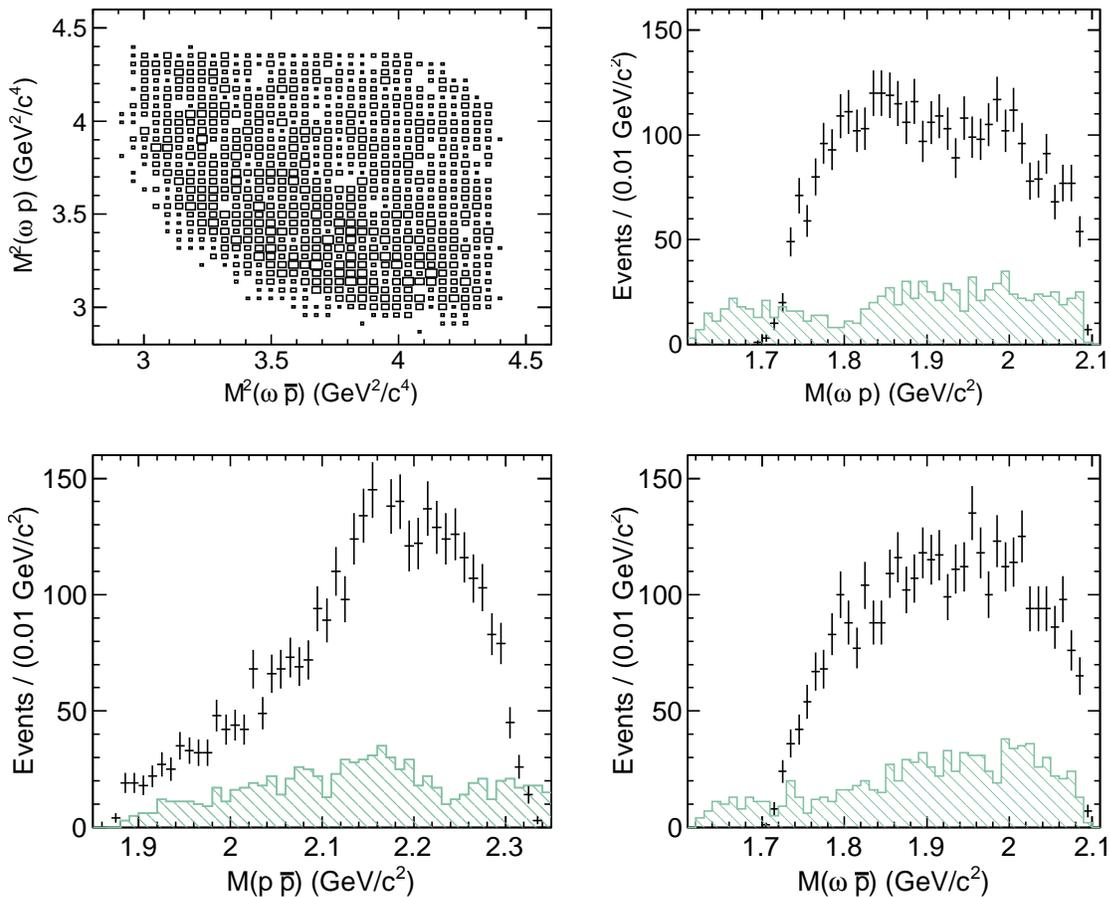}
\caption{Dalitz plot and $p\bar{p}$, $\omega p$, $\omega \bar{p}$ invariant-mass spectra of
$J/\psi \rightarrow \omega p\bar{p}$ candidates. The data points with error
bars are from signal region and the hatched areas are from the sideband region.}
\label{mass_dalitz}
\end{figure}

\end{widetext}

\begin{figure}
\includegraphics[width=8cm]{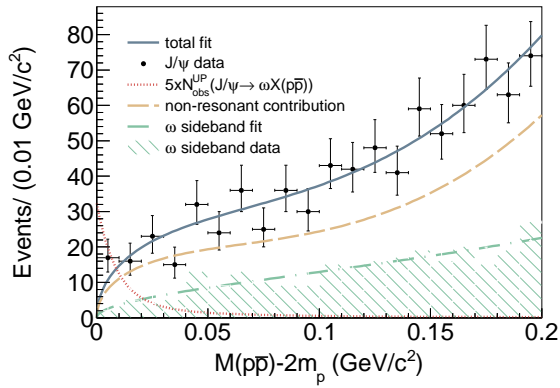}
\caption{Near-threshold $p \bar{p}$ invariant-mass spectrum. The
signal $J/\psi \rightarrow \omega X(p\bar{p})\rightarrow \omega p
\bar{p}$ is described by an acceptance-weighted Breit-Wigner function,
and and signal yield is consistent with zero. The dotted
line is the shape of the signal which is normalized to five times the
estimated upper limit. The dashed line is the non-resonant contribution 
described by the function $f(\delta)$ and the dashed-dotted line is the 
non $\omega p\bar{p}$ contribution which
is estimated from $\omega$ sidebands. The solid line is the total
contribution of the two components. The hatched area is from the sideband region.}
\label{fitmodel}
\end{figure}

Candidate $J/\psi \rightarrow \omega p\bar{p}$ events are selected
with the mass window requirement $0.753\,\text{GeV}/c^{2}<M(\gamma
\pi^{0})<0.813\,\text{GeV}/c^{2}$, and the Dalitz plot of these events
is shown in Fig.~\ref{mass_dalitz}. There are no obvious structures in
the Dalitz plot, though the distribution is different from the pure
$\omega p\bar{p}$ phase space distribution.  The corresponding
$p\bar{p}$, $\omega {p}$ and $\omega \bar{p}$ invariant-mass spectra
are also presented in Fig.~\ref{mass_dalitz}.  The data points with
error bars are from signal region and the hatched area are from the
sideband region. 
the mass threshold is shown in Fig.~\ref{fitmodel}.

To obtain the number of $J/\psi \rightarrow \omega
X(p\bar{p})\rightarrow \omega p \bar{p}$ events, an unbinned maximum
likelihood fit is performed to the $p\bar{p}$ invariant mass around
the mass threshold. In the fit, the spin-parity of $X(p\bar{p})$ is
assumed to be $0^{-}$, and the signal of $X(p\bar{p})$ in the $J/\psi
\rightarrow \omega X(p\bar{p})\rightarrow \omega p \bar{p}$ decay is
parametrized by an acceptance-weighted $\mathcal{S}$-wave
Breit-Wigner function :
\begin{equation}
 BW(M) \simeq \frac{q^{2L+1}k^{3}}{(M^{2}-M^{2}_{0})^{2}+M^{2}_{0}\Gamma^{2}}
\times \varepsilon_{\text{rec}} (M) \ .
\end{equation}

%
Here, $q$ is the momentum of the proton in the $p\bar{p}$ rest frame;
$k$ is the the momentum of the $\omega$ meson; $L=0$ is the relative
orbital angular momentum; $M$ is the invariant mass of $p\bar{p}$;
$M_{0}$ and $\Gamma$ are the mass and width of the $X(p\bar{p})$,
respectively, which are taken from BESIII results~\cite{BAM_00004b};
$\varepsilon_{\text{rec}}$ is the detection efficiency. 
The non-$\omega$ background is presented by a function of the form
$f(\delta)=N(\delta^{1/2}+a_{1}\delta^{3/2} +a_{2}\delta^{5/2})$ with 
$\delta=M_{p\bar{p}}-2m_{p}$ where $m_p$ is the proton mass. The normalization
and shape parameters $a_1$ and $a_2$ are determined by a simultaneous fit
to the $M(p\bar{p})$ in $\omega$ signal region and $\omega$ sideband
region $0.09\,\text{GeV}/c^{2}<|M(\gamma \pi^{0})-0.783|<0.12\,\text{GeV}/c^{2}$. 
The non-resonant $J/\psi \rightarrow \omega p
\bar{p}$ events are also described by the function $f(\delta)$, where
the normalization and shape parameters are allowed to float. The fit
results are shown in Fig.~\ref{fitmodel}, and the number of
$X(p\bar{p})$ events is $0\pm 1.6$.  A Bayesian approach~\cite{pdg}
estimate the upper limit of $\mathcal{B}(J/\psi \rightarrow \omega
X(p\bar{p}) \rightarrow \omega p \bar{p})$, and $N_{\text{obs}}<9$ at
95\% C. L. is determined by finding the value
$N^{\text{UP}}_{\text{obs}}$ with
\begin{equation}
\frac{\int_{0}^{N^{\text{UP}}_{\text{obs}}} \mathcal{L} dN_{obs}} {\int_{0}^{\infty} \mathcal{L} dN_{obs}}=0.95,
\end{equation}
where $N_{obs}$ is the number of signal events, and $\mathcal{L}$ is
the value of the likelihood function with the $N_{obs}$ value fixed in the fit. The 
upper limit on the product of branching fractions is calculated with
\begin{widetext}
\begin{equation}
\mathcal{B}(J/\psi \rightarrow \omega X(p\bar{p})\rightarrow \omega p \bar{p})<
\frac{N^{\text{UL}}_{\text{obs}}}{N_{J/\psi}\times (1-\sigma_{\text{sys.}}) \times \mathcal{B}(\omega \rightarrow \gamma \pi^{0}) \times \mathcal{B}(\pi^{0}\rightarrow \gamma \gamma)\times \varepsilon_{\text{rec}}},
\end{equation}
\end{widetext}
where $\sigma_{sys.}$ is the total systematic uncertainty which will be
described in the next section. The upper limit on the product of
branching fractions is $\mathcal{B}(J/\psi \rightarrow \omega X(p\bar{p})\rightarrow
\omega p \bar{p})<3.9 \times 10^{-6}$ at the 95\% C.L..

An alternative fit with a Breit-Wigner function including the J\"ulich FSI
\begin{equation}
BW(M) \simeq \frac{f_{\text{FSI}} \times q^{2L+1}k^{3}}{(M^{2}-M^{2}_{0})^{2}+M^{2}_{0}\Gamma^{2}}\times \varepsilon_{\text{rec}} (M),
\end{equation}
for $X(p\bar{p})$ is performed. Here, $f_{\text{FSI}}$ is the J\"ulich
FSI correction factor~\cite{Sibirtsev}. The mass and width of
$X(p\bar{p})$ are taken from the previous BESIII PWA
results~\cite{BAM_00004b}. The upper limit on the product of branching
fractions is determined to be $\mathcal{B}(J/\psi \rightarrow \omega
X(p\bar{p})\rightarrow \omega p \bar{p})<3.7 \times 10^{-6}$ at the
95\% C.L..

\section{Systematic Uncertainties}

Several sources of systematic uncertainties are considered in the
measurement of the branching fractions. These include differences
between data and the MC simulation for the tracking algorithm, the
PID, photon detection, the kinematic fit, as well as the fitting
procedure, the branching fraction of the intermediate states and the
total number of $J/\psi$ events.

The systematic uncertainties associated with the tracking efficiency and PID efficiency
have been studied with $J/\psi \rightarrow p\bar{p}\pi^{+}\pi^{-}$ using a technique similar
to that discussed in Ref.~\cite{BAM_00007}. The
difference of tracking efficiencies between data and MC simulation is $2\%$ per charged
track. The systematic uncertainty from PID is 2\% per proton (anti-proton).

The photon detection systematic uncertainty is studied by comparing the photon
efficiency between MC simulation and the control sample $J/\psi
\rightarrow \rho \pi$. The relative efficiency difference is about 1\%
for each photon~\cite{BAM_00002,BAM_00001}.  Here, 3\% is taken as the
systematic error for the efficiency of detecting three photons.  The
uncertainty due to $\pi^{0}$ reconstruction efficiency is taken as 1\%
~\cite{BAM_00001,BAM_00002}.

To estimate the uncertainty associated with the kinematic fit, selected samples of
$J/\psi \rightarrow \Sigma^{+} \bar{\Sigma}^{-}\rightarrow p \pi^{0}\bar{p}\pi^{0}$
events are used. The kinematic fit efficiency is defined as the ratio between the signal yield of
$\Sigma^{+}$ with or without the kinematic fit. The difference of kinematic fit efficiency between
data and MC is 3\%, and is taken as the systematic uncertainty caused by the kinematic fit.

As described above, the yield of $J/\psi\rightarrow\omega p\bar{p}$ is
derived from a fit to the invariant-mass spectrum of $\gamma\pi^0$
pairs. To evaluate the systematic uncertainty associated with the
fitting procedure, the following two aspects are studied (i) {\it
Fitting region}: In the nominal fit, the mass spectrum of
$\gamma\pi^0$ is fitted in the range from 0.663\,GeV/$c^2$ to
0.903\,GeV/$c^2$. Alternative fits within ranges 0.653\,GeV/$c^2$ to
0.913\,GeV/$c^2$ and 0.673\,GeV/$c^2$ to 0.893\,GeV/$c^2$ are
performed, and the difference in the signal yield of 2\% is taken as
the systematic uncertainty associated with the fit interval.  (ii)
{\it Background shape}: To estimate the uncertainty due to the
background parametrization for the branching fraction
$\mathcal{B}(J/\psi \rightarrow \omega p\bar{p})$, a first or third
order instead of a second-order Chebychev polynomial is used in the
fitting.  The difference of $1.2\%$ is used as an estimate of the
systematic uncertainty.

\begin{figure}
\includegraphics[width=8cm]{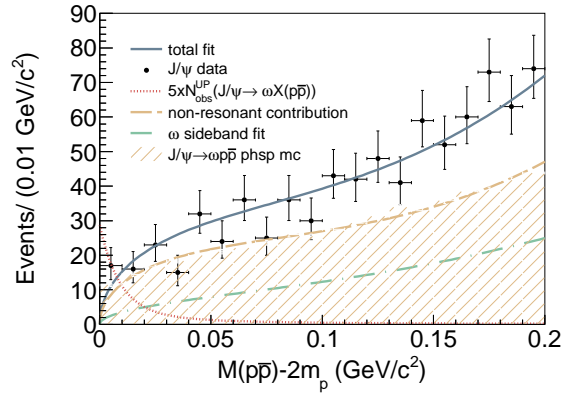}
\caption{Near-threshold $p \bar{p}$ invariant-mass spectrum. The
signal $J/\psi \rightarrow \omega X(p\bar{p})\rightarrow \omega p
\bar{p}$ is described by an acceptance-weighted Breit-Wigner function,
and and signal yield is consistent with zero. The dashed line is the
non-resonant contribution fixed to a phase space MC simulation
of $J/\psi \rightarrow \omega  p \bar{p}$ and the dashed-dotted line 
is the non $\omega p\bar{p}$ contribution which
is estimated from $\omega$ sidebands. The solid line is the total
contribution of the two components. The hatched area is from  
a phase space MC simulation of $J/\psi \rightarrow \omega p \bar{p}$.}
\label{phsp_fixed}
\end{figure}

For the upper limit on the branching fraction $\mathcal{B}(J/\psi
\rightarrow \omega X(p\bar{p})\rightarrow \omega p \bar{p})$, the
systematic uncertainty associated with the fitting procedure is
estimated by fixing the shape of the non-resonant 
contribution to a phase space MC simulation of $J/\psi \rightarrow \omega p \bar{p}$, 
which is presented by Figure.~\ref{phsp_fixed};
enlarging/reducing the normalization of the non-$\omega$ contribution
 by 7\% (the difference of the estimation 
of non-$\omega$ background level between data and inclusive MC);
and varying the sideband region to
$0.095\,\text{GeV}/c^{2}<|M(\gamma \pi^{0})-0.783|<0.115\,\text{GeV}/c^{2}$
and $0.085\,\text{GeV}/c^{2}<|M(\gamma \pi^{0})-0.783|<0.125\,\text{GeV}/c^{2}$. 
When fitting with or without the FSI effect, the signal yields for the 
alternative fits are lower or equal to the nominal fit, therefore the conservative 
upper limit from the fit without FSI correction is reported.

Various distributions obtained with data and the phase-space MC sample
have been compared and some discrepancies are observed. To determine
the systematic error on the detection efficiency associated with these
discrepancies, an alternative detection efficiency is estimated by the
re-weighting phase-space MC samples. The difference in detection
efficiency compared to the nominal one is 7\% and taken as a
systematic uncertainty.  The number of $J/\psi$ events is determined
from an inclusive analysis of $J/\psi$ hadronic events and an
uncertainty of 1.24\% is associated to it~\cite{BAM_00011}.  The
uncertainties due to the branching fractions of $\omega \rightarrow
\gamma \pi^{0}$ and $\pi^0 \to \gamma\gamma$ are taken from the
PDG~\cite{pdg}.

\begin{widetext}
\begin{center}

\begin{table}
\caption{Summary of systematic uncertainties. '-' means the corresponding systematic uncertainty is
  negligible.
}
\begin{ruledtabular}
\begin{tabular}{lccc}
 &&Upper limit of&Upper limit of \\
  Source  & $\mathcal{B}(J/\psi\rightarrow \omega p\bar{p})$ &  $\mathcal{B}(J/\psi \rightarrow \omega X(p\bar{p})\rightarrow \omega p \bar{p})$ &  $\mathcal{B}(J/\psi \rightarrow \omega X(p\bar{p})\rightarrow \omega p \bar{p})$ with FSI \\

\hline
Tracking& 4\%   & 4\% & 4\% \\
PID     & 4\%   & 4\% & 4\% \\
Photon  & 3\%   & 3\% & 3\% \\
Kinematic Fit& 3\% & 3\%& 3\%\\
$\pi^{0}$ reconstruction &1\% & 1\% &1\%\\
Fitting region &2\%& $-$ & $-$ \\
Background Shape& 1\% & $-$ & $-$\\
Branching fraction of intermediate state & 3\% & 3\% & 3\%\\
Total $J/\psi$ numbers & 1.24\% & 1.24\% &1.24\%\\
MC Generator  & 7\% & $-$ &$-$\\
\hline \
Total uncertainty& 10.3\% & 7.8\% &7.8\% \\
\end{tabular}
\end{ruledtabular}

\end{table}
\end{center}
\end{widetext}

 \section{Summary}
In summary, using $(225.3 \pm 2.8)\times 10^{6}$ $J/\psi$ events
collected with the BESIII detector, the decay of $J/\psi \rightarrow
\omega p\bar{p}$ in the decay mode $\omega \rightarrow \gamma \pi^{0}$
is studied. The branching fraction $\mathcal{B}(J/\psi\rightarrow
\omega p\bar{p})$ is measured to be $(9.0\pm0.2 \, (\text{stat.})\pm
0.9 \,(\text{syst.}))\times 10^{-4} $.  No obvious enhancement around
the $p\bar{p}$ invariant-mass threshold is observed. At the 95\% C.L.,
the upper limits on the product of branching fractions
$\mathcal{B}(J/\psi\rightarrow \omega X(p\bar{p}) \rightarrow \omega
p\bar{p})$ are measured to be $3.7 \times 10^{-6}$ and $3.9 \times
10^{-6}$ with and without accounting for the J\"ulich FSI effect,
respectively.  As isospin for $J/\psi\rightarrow \gamma p\bar{p}$ and
$\omega p\bar{p}$ should both favor $I=0$ ($I=1$ should be suppressed
in $J/\psi \rightarrow \gamma p\bar{p}$ as in other $J/\psi$ radiative
decays), the non-observation of $X(p\bar{p})$ in $\omega p\bar{p}$
disfavors the pure FSI interpretation for the $p\bar{p}$ threshold
enhancement in the decay $J/\psi \rightarrow \gamma p\bar{p}$.

 \section{Acknowledgment}
The BESIII collaboration thanks the staff of BEPCII and the computing center for their hard efforts. This work is supported in part by the Ministry of Science and Technology of China under Contract No. 2009CB825200; National Natural Science Foundation of China (NSFC) under Contracts Nos. 10625524, 10821063, 10825524, 10835001, 10935007, 11125525, 11235011; Joint Funds of the National Natural Science Foundation of China under Contracts Nos. 11079008, 11179007; the Chinese Academy of Sciences (CAS) Large-Scale Scientific Facility Program; CAS under Contracts Nos. KJCX2-YW-N29, KJCX2-YW-N45; 100 Talents Program of CAS; German Research Foundation DFG under Contract No. Collaborative Research Center CRC-1044; Istituto Nazionale di Fisica Nucleare, Italy; Ministry of Development of Turkey under Contract No. DPT2006K-120470; U. S. Department of Energy under Contracts Nos. DE-FG02-04ER41291, DE-FG02-05ER41374, DE-FG02-94ER40823; U.S. National Science Foundation; University of Groningen (RuG) and the Helmholtzzentrum fuer Schwerionenforschung GmbH (GSI), Darmstadt; WCU Program of National Research Foundation of Korea under Contract No. R32-2008-000-10155-0.

  \end{document}